\documentstyle[twoside,fleqn,espcrc2,epsfig]{article}

\newcommand{\AmS}{{\protect\the\textfont2
  A\kern-.1667em\lower.5ex\hbox{M}\kern-.125emS}}

\font\sfhuge= cmss24  at 20truept
\font\sfLARGE= cmss14  at 14truept
  at 12truept
\font\sfmed= cmss10  at 10truept
\font\sfsml= cmss8   at  8truept

\title{Determinations of $\alpha_s$ using
  JADE data of e$^+$e$^-$ Annihilations \mbox{at $\sqrt{s} = 22$ to $44$~GeV} 
}

\author{P.A.~Movilla~Fern\'andez\address{III. Physikalisches Institut A,
        RWTH Aachen --- Physikzentrum, \\ 
        Sommerfeldstr., D-53056 Aachen, Germany}%
       }
       
\begin{document}
\addtolength{\textheight}{-68mm}
\begin{titlepage}
\thispagestyle{empty}

\vspace*{-10mm}
\vbox to 245mm{

\hbox to \textwidth{ \hsize=\textwidth
\vbox{
\hbox {
\epsfig{file=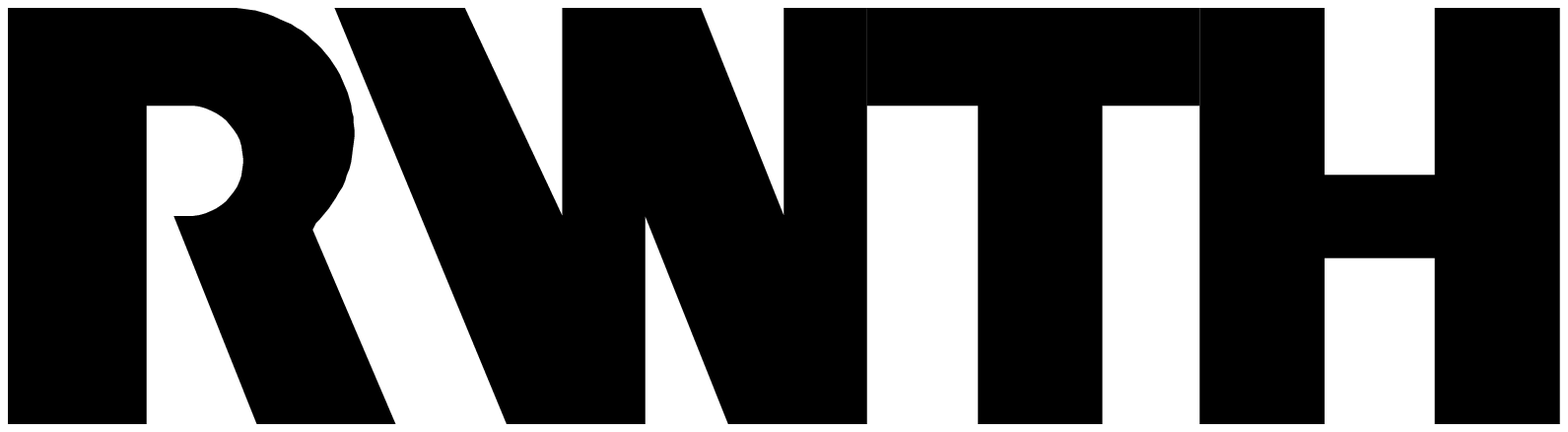,height=20mm}
} 
}
\vbox{
{
\hbox{\sfmed RHEINISCH-\hss}\vspace*{+0.150mm}
\hbox{\sfmed WESTF\"ALISCHE-\hss}\vspace*{+0.150mm}
\hbox{\sfmed TECHNISCHE-\hss}\vspace*{+0.150mm}
\hbox{\sfmed HOCHSCHULE-\hss}\vspace*{+0.150mm}
\hbox{\sfmed AACHEN\hss}\vspace*{0.200mm}
}
}
\vbox{ \hsize=58mm 
{
\hspace*{0pt\hfill}\hbox{\sfLARGE\hspace*{0pt\hfill}        PITHA 97/33\hss}\vspace*{-2mm}
\hspace*{0pt\hfill}\hbox{        \hspace*{0pt\hfill} \rule{45mm}{1.0mm}\hss}
\hspace*{0pt\hfill}\hbox{\sfLARGE\hspace*{0pt\hfill}       August 1997\hss}\vspace*{2.3mm}
}
}
}

\vspace*{5cm}

\begin{center}
{\huge\bf
Determinations of $\alpha_s$ \\[2mm] using
  JADE data of e$^+$e$^-$ Annihilations \\[2mm] \mbox{at $\sqrt{s} = 22$ to $44$~GeV}
}
\end{center}
\vspace*{2cm}
\begin{center}
\Large
Pedro~A.~Movilla~Fern\'andez \\
\bigskip 
\bigskip
III. Physikalisches Institut, Technische Hochschule Aachen\\
D-52056 Aachen, Germany
\end{center}

\vspace*{0pt\vfill}
\vfill

\vspace*{-5mm}
\noindent
\hspace*{-5mm}
\hbox {
\rule{\textwidth}{0.3mm}
}

\vspace*{3mm}
\noindent
\begin{minipage}{\textwidth}

\vbox {\vsize=60mm
\hbox to \textwidth{\hsize=\wd0
\hbox {\hspace*{-5mm}

\vbox{ 
\hbox to \textwidth{\hss\sfhuge PHYSIKALISCHE INSTITUTE\hss }\vspace*{2.0mm}
\hbox to \textwidth{\hss\sfhuge      RWTH AACHEN\hss }\vspace*{2.0mm}
\hbox to \textwidth{\hss\sfhuge    Physikzentrum\hss }\vspace*{2.0mm}
\hbox to \textwidth{\hss\sfhuge D-52056 AACHEN, GERMANY\hss}
}

}
}
}

\end{minipage}
}

\end{titlepage}

\setlength{\textheight}{202mm}
\setlength{\textwidth}{160mm}
\setlength{\oddsidemargin}{-4mm}
\setlength{\evensidemargin}{4mm}
\setlength{\topmargin}{16mm}
\setlength{\headheight}{13mm}
\setlength{\headsep}{21pt}
\setlength{\footskip}{30pt}
\clearpage

\begin{abstract}
\vspace*{-52mm}
\hbox{\sfsml Talk presented at the {\em QCD'97}, 
             Montpellier, France, July 3-9, 1997.
}
\vspace*{48mm}
  Data recorded by the JADE experiment at the PETRA e$^+$e$^-$
  collider were used to measure distributions of new event shape
  observables.  The distributions were compared with resummed QCD
  calulations (${\cal O}(\alpha_s^2)$+NLLA), and the strong coupling
  constant $\alpha_s$ was determined at $\sqrt{s}=$ 22, 35 and 44~GeV.
  The results are in agreement with previous combined results of PETRA
  but have smaller uncertainties.  Together with corresponding data
  from LEP, the energy dependence of $\alpha_s$ is significantly
  tested and is found to be in good agreement with the QCD
  expectation.
\end{abstract}

\maketitle

\section{INTRODUCTION}
Summaries of measurements of $\alpha_s$ from various processes and at
different energy scales $Q$ demonstrate \cite{bib-world-alphas-sb}
that the energy dependence of $\alpha_s$ is in good agreement with the
prediction of Quantum Chromodynamics (QCD).  High precision tests are
provided by e$^+$e$^-$ annihilation data which were collected between
some GeV and nearly 200 GeV.  Unfortunately, in the interesting
intermediate energy region where $\alpha_s$ runs very quickly, the
``best'' value of $\alpha_s$, 0.14 $\pm$
0.2~\cite{bib-world-alphas-sb}, is a combined result at $\sqrt{s}=$
35~GeV based on measurements at PETRA that are affected by large
theoretical uncertainties because only QCD predictions up to ${\cal
  O}(\alpha_{s}^{2})$ were available for event shape observables.

Significant progress has been made in perturbative QCD calculations
since 1992. Observables have been proposed for which perturbative
predictions are extended beyond the next-to-leading-order through the
inclusion of leading and next-to-leading logarithms~\cite{bib-NLLA}
which are summed to all orders of $\alpha_s$ (NLLA).  Further, the
precise measurements of the hadronisation process have improved the
modelling of hadronic final states with Monte Carlo programs.

It is desirable to apply these LEP-established developments
consistently to lower energy data in order to improve quantitative
studies of the running of $\alpha_s(Q)$ over large ranges of the
energy scale $Q$, using identical experimental techniques and
theoretical calculations in order to minimise point-to-point
systematic uncertainties.  In the following, ${\cal
  O}(\alpha_s^2)$+NLLA determinations of $\alpha_s$ using data from
the JADE experiment at PETRA at $\sqrt{s} = 22$, $35$, and $44$ GeV
are presented \cite{bib-fernandez}.

\section{JADE DATA}
The JADE detector~\cite{bib-JADEdet} was one of the five experiments
at the PETRA electron-positron accelerator.  It has operated from 1979
until 1986 at centre-of-mass energies of $\sqrt{s} = 12$ to $46.7$
GeV.  The main components of the detector were the central jet chamber
to measure charged particle tracks and the lead glass calorimeter to
measure electromagnetic showers.

Multihadronic events were selected by the standard JADE selection
cuts~\cite{bib-JADEdet} which reduce the background from
$\gamma\gamma$ and $\tau$-pair events to less than $0.1\%$ and $1\%$,
respectively~\cite{bib-JADEeventsel}.  The final numbers of events
which were retained for this analysis are listed in
Table~\ref{tab-eventnumbers}, together with the corresponding
integrated luminosities.
\begin{table}[t]
\caption{\label{tab-eventnumbers} \small 
Number of multihadronic events in JADE data   
   and corresponding integrated luminosities.
}
\vspace*{-2mm}
\begin{center}
\begin{tabular}{|c|c||c|c|}
\hline
year & $\sqrt{s}\ [\mathrm{GeV}]$ & data & luminosity \\ 
\hline\hline
   1981 &  $22$      &             $\ 1404$ &  2.4 $pb^{-1}$     \\ \hline
1984/85 &  $39.5$-$47.8$ &         $\ 6158$ &  80 \   $pb^{-1}$      \\ \hline
   1986 &  $35$      &   $20\thinspace 926$ &  40 \  $pb^{-1}$      \\
\hline
\end{tabular}
\end{center}
\vspace*{-8mm}
\end{table}
The physical results of previous publications by JADE can be very well
reproduced~\cite{bib-fernandez}.

Monte Carlo detector simulation corresponding to the JADE data were
retrieved for $35$ and $44$~GeV.  They were generated using the JETSET
QCD shower event generator version 6.3~\cite{bib-JETSET}.  In general,
there is good agreement of the detector simulation with data for all
event shape distributions studied here.  The simulated data can thus
be used to correct for detector effects in the measured data.

\section{DETERMINATION OF $\alpha_s$}

\subsection{Event shapes and jetrates}
 
From the data samples, the event shape distributions of thrust $T$,
the heavy jet mass $M_H$, the total and wide jet broadening $B_T$ and
$B_W$ and the differential 2-jet event rate using the Durham jet
finder $y_{23}$ were determined(c.f.~\cite{bib-fernandez}).  For the
latter three observables, no experimental results have previously been
presented at these energies.

\begin{figure}[!t]
\vspace*{-3mm}
\centerline{
\epsfig{file=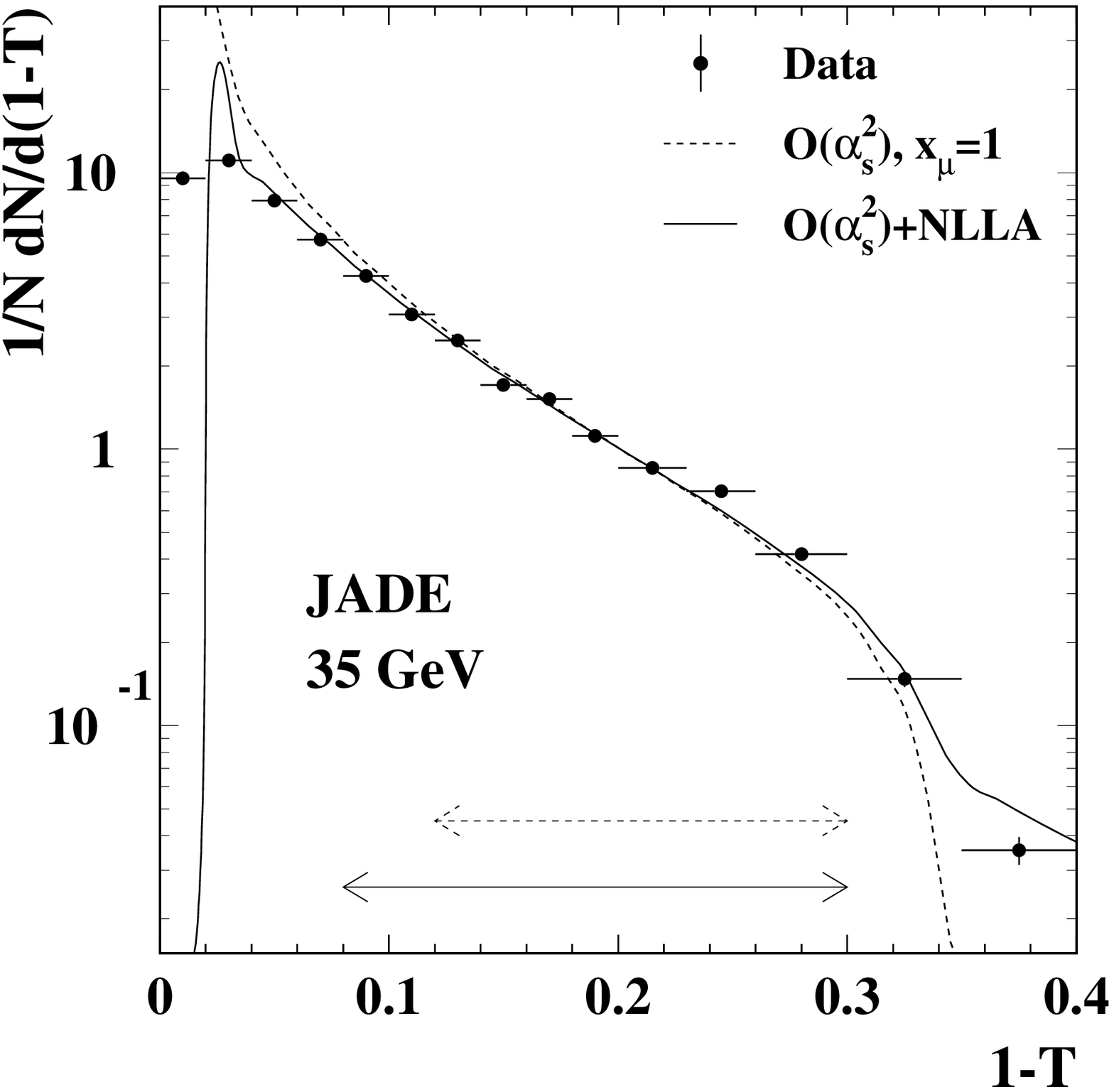,width=.35\textwidth,clip=}
}
\vspace*{-5.mm}
\centerline{
\epsfig{file=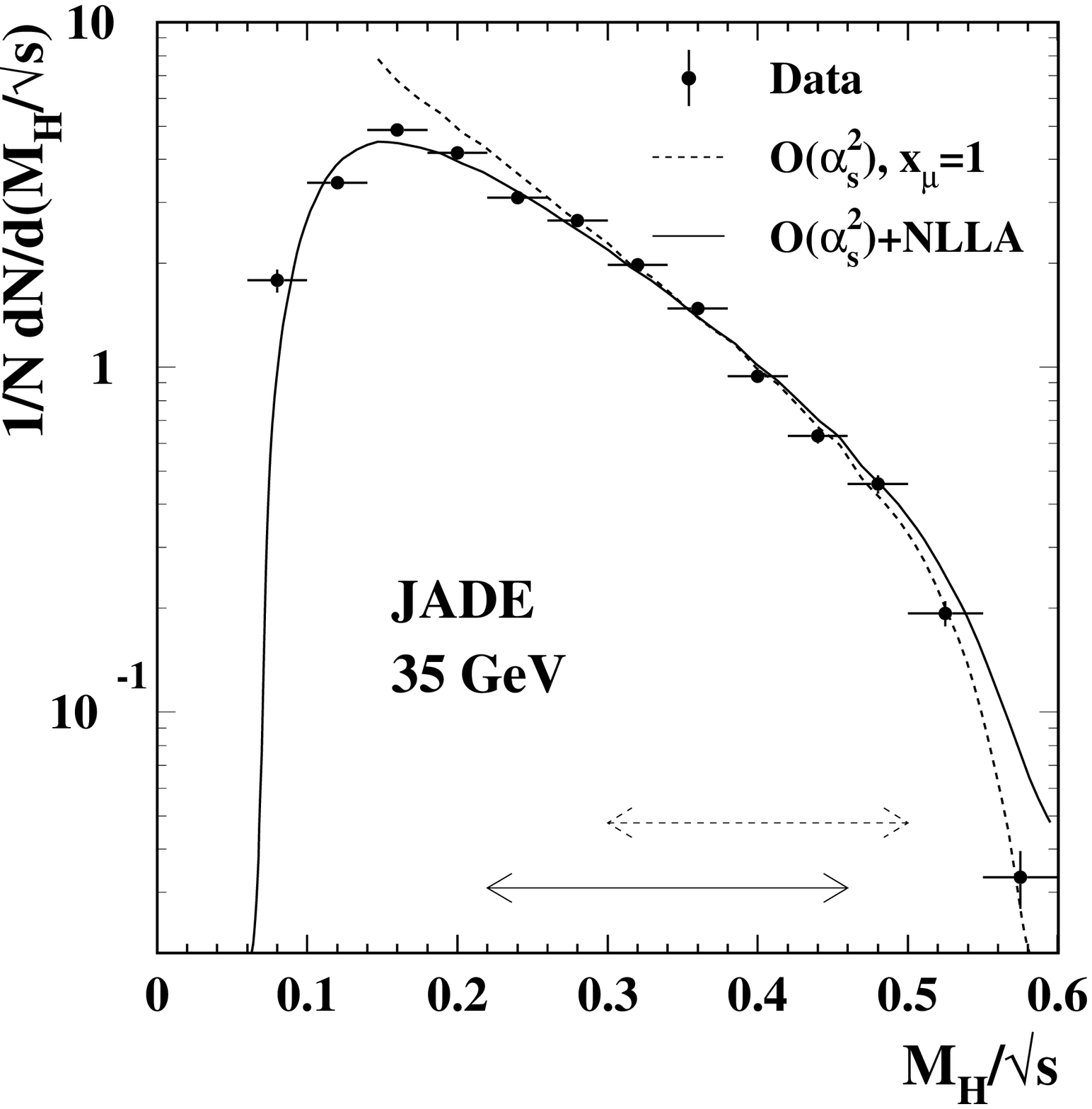,width=.35\textwidth,clip=}
}
\vspace*{-11mm}
\caption{\label{fig-asresult-1} \small
  Distributions for $1-T$ and $M_H/\protect\sqrt{s}$ at
  $\protect\sqrt{s} = 35$~GeV corrected to parton level. The fits of
  the QCD predictions are overlaid and the fit ranges are indicated by
  the arrows. }
\vspace*{-4mm}
\end{figure}

\subsection{Correction procedure}

The event shape data were corrected for the limited acceptance and
resolution of the detector and for initial state photon radiation
effects by applying a bin-by-bin correction procedure.  It turns out
that there is an excellent agreement between the data and the model
prediction at {\em hadron level} over the whole kinematic range of the
observables even for those ones which were never tested at such low
energies \cite{bib-fernandez}.

No detector simulation were available for the $22$~GeV data.
Therefore only the differential 2-jet rate was considered because it
is known to depend to a lesser extend on detector effects.  The 35~GeV
correction was applied to the 22~GeV data since the correction factors
for this observable were found to show no significant energy
dependence~\cite{bib-fernandez}.

In a second step, the data distributions were corrected for
hadronisation effects by applying bin-by-bin correction factors
derived from the ratio of the JETSET generated distributions before
and after hadronisation.  The data distributions, thus corrected to
the {\em parton level}, can be compared directly to analytic QCD
calculations.

\subsection{${\cal O}(\alpha_s^2)$+NLLA predictions}

\begin{figure}[!t]
\vspace*{-2.mm}
\centerline{
\epsfig{file=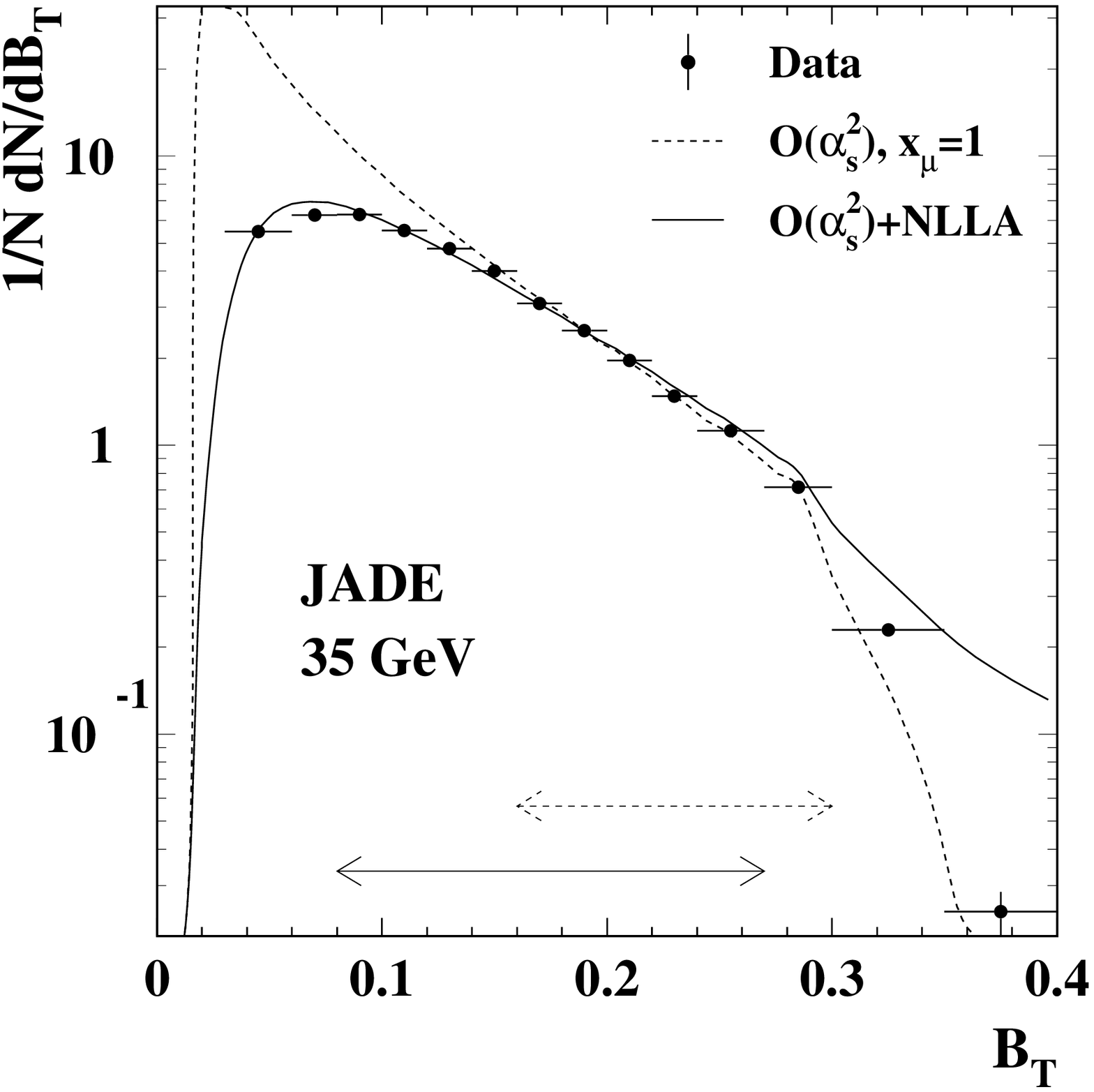,width=.35\textwidth,clip=}
}
\vspace*{-5.mm}
\centerline{
\epsfig{file=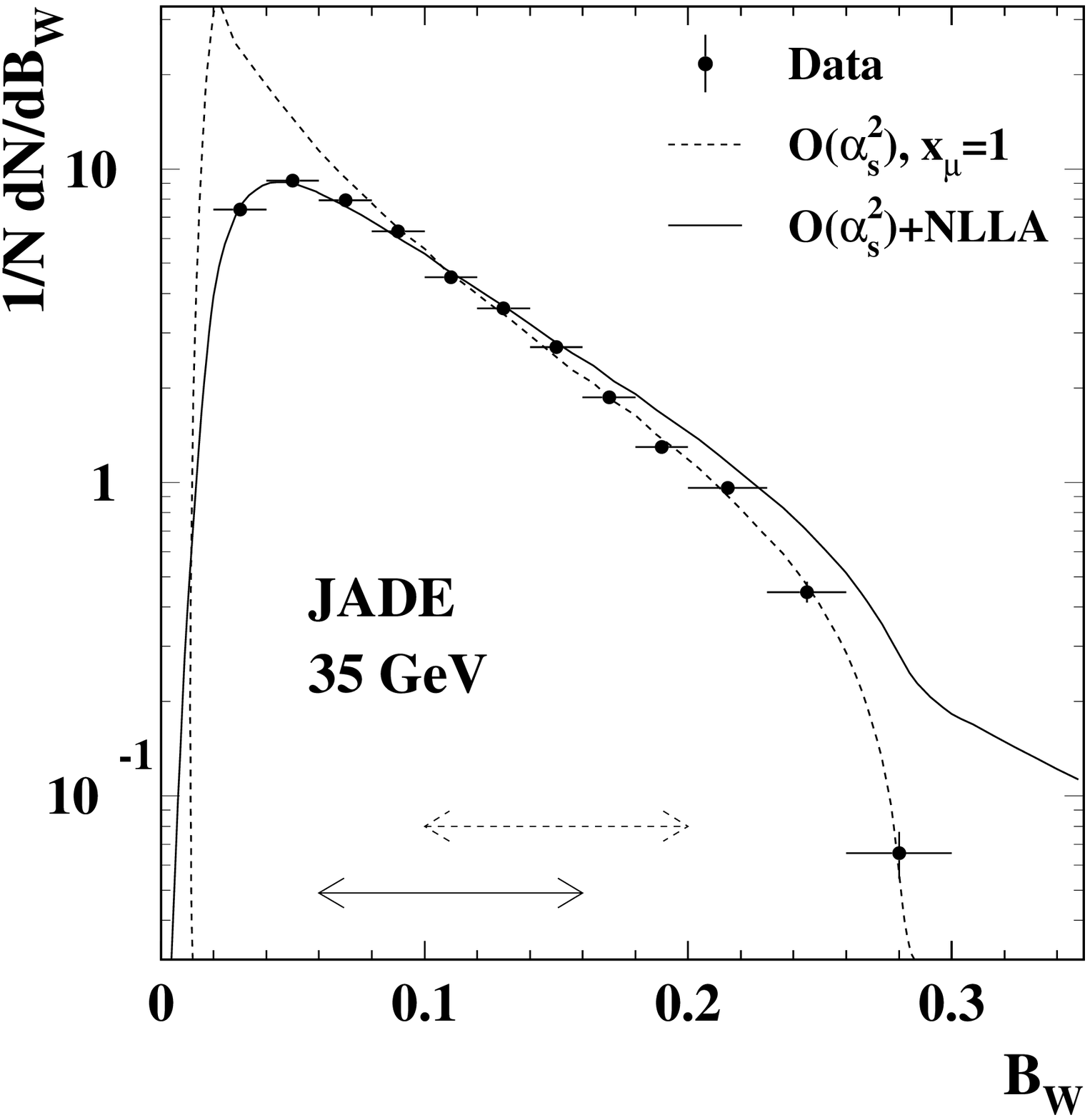,width=.35\textwidth,clip=}
}
\vspace*{-11mm}
\caption{\label{fig-asresult-2} \small
  Distributions for $B_T$ and $B_W$ at $\protect\sqrt{s} =
  35$~GeV corrected to parton level. The fits of the
  QCD predictions
  are overlaid and the fit ranges are indicated by the arrows. } 
\vspace*{-4mm}
\end{figure}

\begin{figure}[!t]
\vspace*{-2.mm}
\centerline{
\epsfig{file=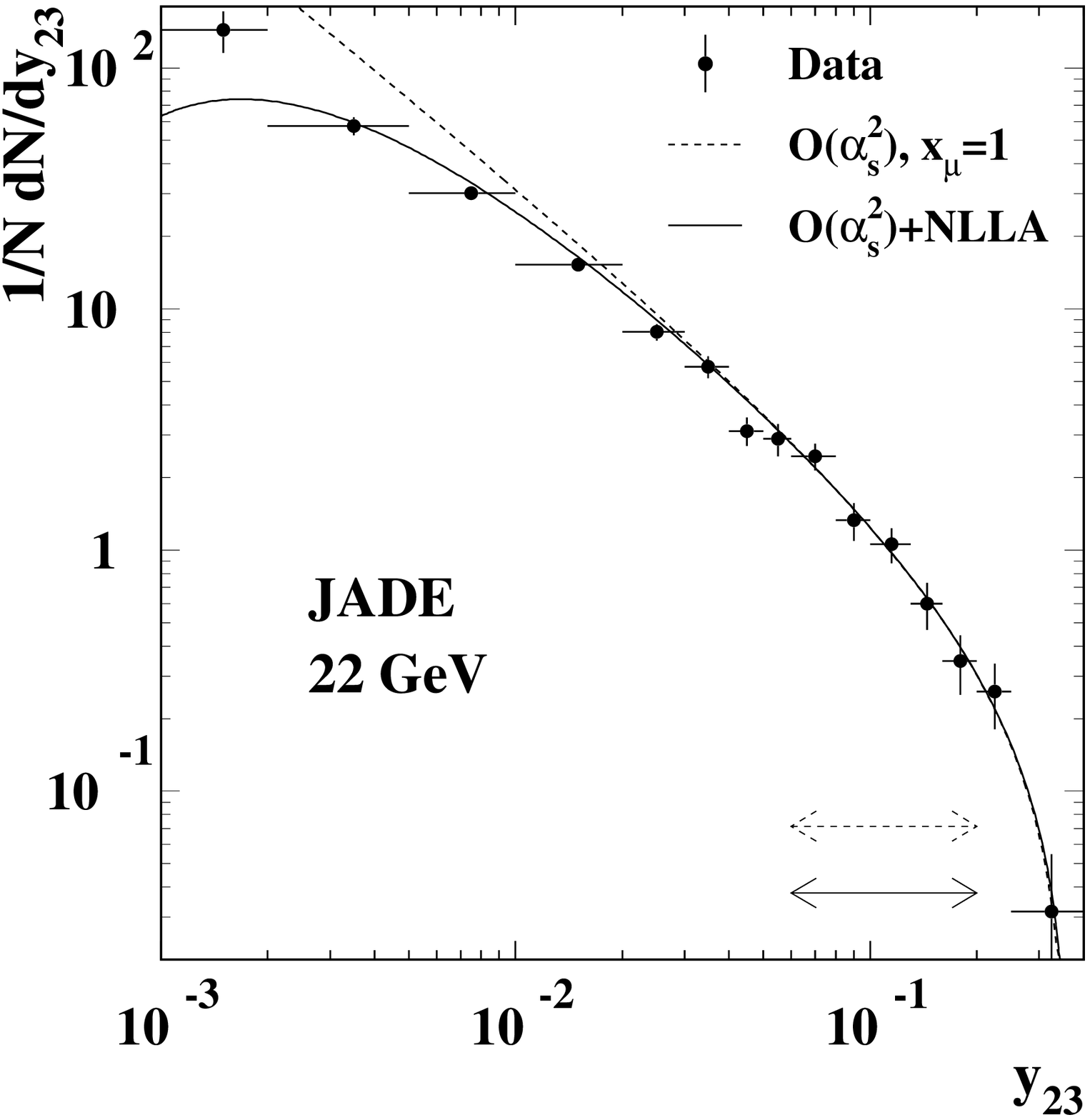,width=.35\textwidth,clip=}
}
\vspace*{-5mm}
\centerline{
\epsfig{file=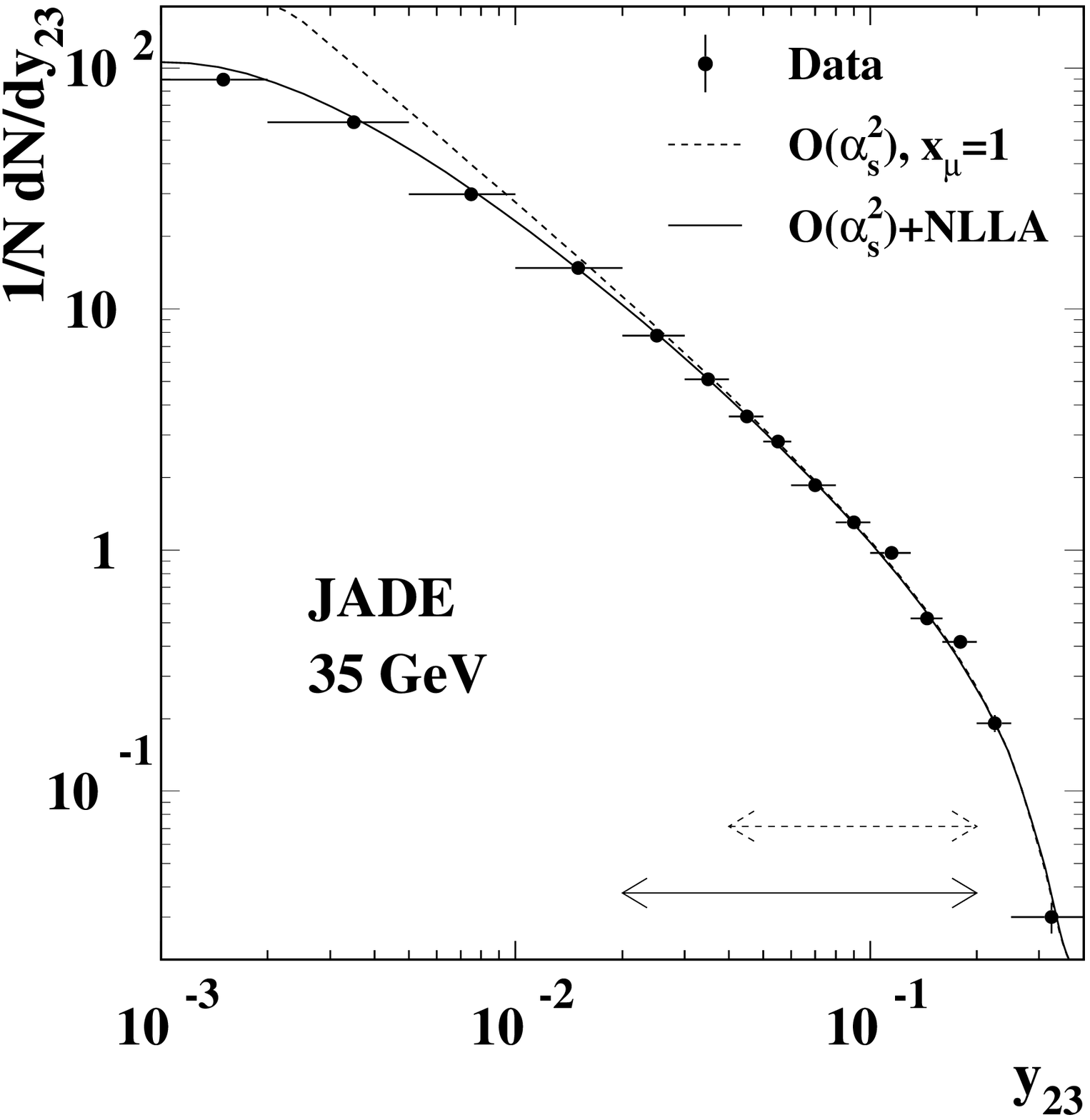,width=.35\textwidth,clip=}
}
\vspace*{-11mm}
\caption{\label{fig-asresult-3} \small
  Distributions for $y_{23}$ at $\protect\sqrt{s} = 22$ and 35~GeV corrected
  to parton level. The fits of the QCD predictions are overlaid and
  the fit ranges are indicated by the arrows. }  
\vspace*{-4mm}
\end{figure}

The event shape observables used in the analysis are predicted in
perturbative QCD by a combination of the ${\cal
  O}(\alpha_s^2)$~\cite{bib-ERT} and the NLLA~\cite{bib-NLLA}
calculations.  The ${\cal O}(\alpha_s^2)$ calculation yields an
expression of the form
\begin{displaymath}
    R_{{\cal O}\alpha_s^2}(y) = 1 + A(y)\left(\frac{\alpha_s^2}{2\pi}\right) 
                    + B(y)\left(\frac{\alpha_s^2}{2\pi}\right)^2,
\end{displaymath}
where 
$R(y) = \int_{0}^{y} {{\mathrm{d}} y}\ 1/\sigma_0 \cdot 
        {{\mathrm{d}}}\sigma/{{\mathrm{d}}y}$
is the cumulative cross-section of an event shape 
observable $y$
normalised to the lowest order Born cross-section $\sigma_0$. 
The NLLA calculations
give an expression for $R(y)$ in the form:
\begin{eqnarray*}
R_{\mathrm{NLLA}}(y) = \left(1 + C_1\left(\frac{\alpha_s^2}{2\pi}\right) 
                               + C_2\left(\frac{\alpha_s^2}{2\pi}\right) ^2\right) 
                 \cdot      \\  \cdot
                       \exp\left[L\, g_1\!\left(\frac{\alpha_s^2}{2\pi} L\right)
                                 +   g_2\!\left(\frac{\alpha_s^2}{2\pi} L\right)\right]
\end{eqnarray*}
where $L = {\mathrm{ln}}(1/y)$. The functions $g_1$ and $g_2$ are given by
the NLLA calculations. The coefficients $C_1$ and $C_2$ are known from the
${\cal O}(\alpha_s^2)$ matrix elements.

The strong coupling $\alpha_s$ was determined by $\chi^2$ fits to the
event shape distributions corrected to the {\em parton level}.  For
the sake of direct comparison with other published results this
analysis closely followed the procedures described in
\cite{bib-OPALresummed}. The so-called ln($R$)-matching scheme was
chosed to merge the ${\cal O}$($\alpha_s^2$) with the NLLA
calculations.  The renormalisation scale factor, $x_{\mu}$ $\equiv
\mu/\sqrt{s}$, was set to $x_{\mu} = 1$ for the main result.

The fit ranges are determined by the requirements of moderate
hadronisation uncertainties within the fit ranges, good
$\chi^2/\mathrm{d.o.f.}$ and reasonably stable fits under changes of
the range.

\subsection{Systematics}
To study systematic uncertainties~\cite{bib-fernandez} details of the
event selection and of the correction procedure were modified. The
impact of the hadronisation model of the JETSET generator was studied
by varying several significant model parameters. Mass effects are
estimated conservatively by excluding the b-quark from the correction
procedure.  Uncertainties due to contributions of unknown higher order
corrections are estimated by the changes of $\alpha_s$ when varying
the renormalisation scale factor $x_{\mu}$ in the range of 0.5 to 2.0.
Any resulting deviation from the main result was considered as
systematic error.
\subsection{Results}

As an example, the theoretical predictions for the observables $1-T$,
$M_H$, $B_T$ and $B_W$ at $\sqrt{s}=35$ and $y_{23}$ at $\sqrt{s}=22$
and 35~GeV are presented in Figures \ref{fig-asresult-1} -
\ref{fig-asresult-3} and superimposed on the corresponding measured
distributions corrected to {\em parton level}.  In general, resummed
calculations (solid lines) give a good description of the observables
over their entire range of values even at the low PETRA energies,
while the ${\cal O}(\alpha_s^2)$ calculations (dashed lines) which are
shown for comparison fail in describing the 2-jet region.

\begin{figure}[!t]
\vspace*{-10mm}
\centerline{
\epsfig{file=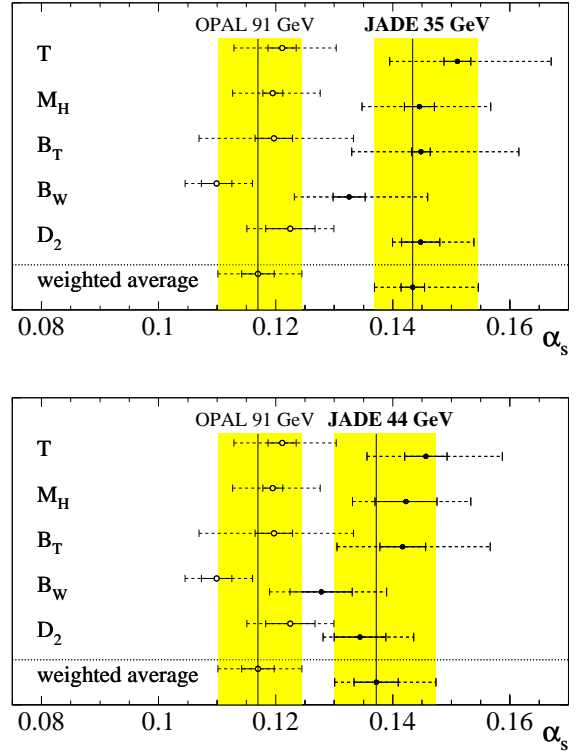,width=.56\textwidth,clip=}
}
\vspace*{-14mm}
\caption{\label{fig-asresult-numbers} \small
  Values of $\alpha_s$(35~GeV) and $\alpha_s$(44~GeV) derived from
  ${\cal O}(\alpha_s)$+NLLA fits to event shape distributions,
  compared with OPAL results. The experimental and statistical
  uncertainties are represented by the solid, the total error by the
  dashed error bars.  } \vspace*{-6mm}
\end{figure}

The values of $\alpha_s$ and the errors obtained at $35$ and $44$~GeV
are shown in Figure~\ref{fig-asresult-numbers} in comparison with the
$\alpha_s$ values measured by OPAL at $\sqrt{s} =
M_{Z^0}$~\cite{bib-OPALresummed}.  The values of $\alpha_s$ exhibit a
similar scattering pattern at all energies.  This demonstrates the
strong correlation of the systematic uncertainties at JADE and OPAL.

The individual results of the five observables at each energy were
combined~\cite{bib-fernandez} into a single value.  The final results
for $\alpha_s$ are
\begin{eqnarray*} 
\alpha_s(44{\mathrm{GeV}})= 0.1372 \pm
  0.0017{\mathrm{(stat.)}}  ^{+0.0101}
  _{-0.0069}{\mathrm{(syst.)}} \\
\alpha_s(35{\mathrm{GeV}}) = 0.1434 \pm 0.0010{\mathrm{(stat.)}}
  ^{+0.0112}
  _{-0.0065}{\mathrm{(syst.)}} \\
\alpha_s(22{\mathrm{GeV}}) = 0.1608 \pm 0.0083{\mathrm{(stat.)}}
  ^{+0.0137} _{-0.0058}{\mathrm{(syst.)}}
\end{eqnarray*}
where the result at $22$~GeV is based on $y_{23}$ only.  The main
contributions to the systematic errors come from renormalisation scale
uncertainties and mass effects which causes systematically higher
values for $\alpha_s$.

\section{SUMMARY AND CONCLUSIONS}

JADE data at $\sqrt{s}=$ $22$, $35$ and $44$~GeV were used to measure
$\alpha_s$ simultaneously with old ($T$, $M_H$) and new ($B_T$, $B_W$,
$y_{23}$) event shape observables. The corresponding ${\cal
  O}(\alpha_s^2)$+NLLA predictions are in good agreement with the
measured distributions corrected to parton level even at PETRA
energies.

\begin{figure}[!t]
\vspace*{-5mm}
\centerline{\hspace*{5mm}
\epsfig{file=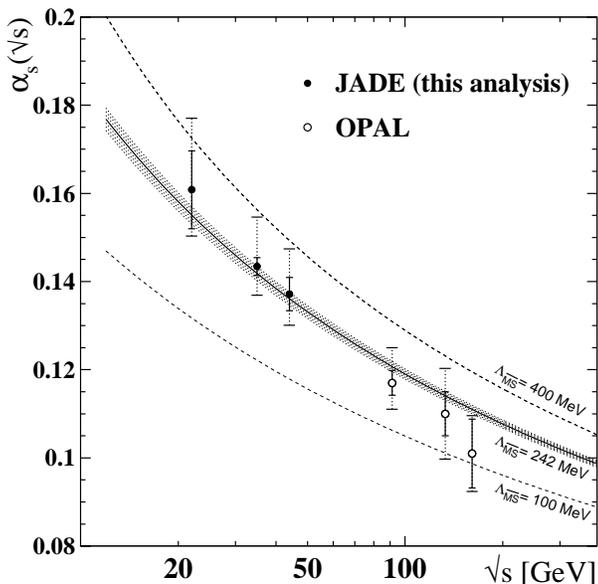,width=.55\textwidth,clip=}
}
\vspace*{-13.2mm}
\caption{\label{fig-world-NLLA-as} \small
  Values of $\alpha_s$ from ${\cal O}(\alpha_s^2)$+NLLA fits at JADE
  and OPAL. The solid error bars are the statistical and experimental
  uncertainties, the dotted error bars are the total errors.  The
  solid line and the shaded band represent a QCD-fit to the data and
  its statistical error, respectively.  } \vspace*{-6.3mm}
\end{figure}

It can be pointed out that the present measurements of $\alpha_s$ are
the most precise in this energy region of the e$^+$e$^-$ continuum.
The individual results are {\em consistent} with each other.  The
errors are dominated by effects of missing higher orders in the QCD
calculations.

Evolving the measurements to $\sqrt{s} = M_{Z^0}$ the results obtained
at $44$, $35$ and $22$~GeV transform to a combined value of
$\alpha_s(M_{Z^0})= 0.122^{+0.008}_{-0.005}$ which is consistent with
the direct measurement at $\sqrt{s}=M_{Z^0}$ by OPAL of $\alpha_s(
M_{Z^0} ) = 0.117^{+0.008}_{-0.006}$\cite{bib-OPALresummed}, for the
same subset of observables.

Similarities between the main detector components of
JADE~\cite{bib-JADEdet} and OPAL~\cite{bib-OPALdet}, as well as
between this analysis and studies performed by
OPAL~\cite{bib-OPALresummed,bib-OPALNLLA} at $\sqrt{s} = 91.2$, $133$,
and $161$~GeV, allow for a precision test of the energy dependence of
$\alpha_s$ between $\sqrt{s} = 22$-$161$~GeV.  The $\alpha_s$
\mbox{values} from OPAL and from this analysis are shown in
Figure~\ref{fig-world-NLLA-as}.  The result of a $\chi^2$ fit of the
${\cal O}(\alpha_s^3)$ QCD prediction to the data is shown by the
solid line corresponding to $\alpha_s(M_{Z^0}) = 0.1207 \pm 0.0012$
and $\chi^2/\mathrm{d.o.f.}$ = 4.9/5 , taking into account only
statistical and experimental uncertainties.

A fit for the hypothesis $\alpha_s=$ const. gives
a $\chi^2/\mathrm{d.o.f.} = 101/5$  which has a vanishing
probability.  The energy dependence of $\alpha_s$ is therefore significantly
demonstrated by the results from the combined JADE and OPAL data.

\section*{QUESTIONS}
\begin{raggedright}
{\em D.~Dissertori, CERN:}  \\
\end{raggedright}
Why did you use only the ln($R$)-matching scheme? Typically
a set of different matching schemes is used, and this
in the end could increase your systematic error.

\vspace*{1ex}
\begin{raggedright}
  {\em P.A.M.~Fern\'andez:}  \\
\end{raggedright}
In order to compare the present measurements consistently with the
results given by OPAL, I applied only the ln($R$)-matching scheme
because it is commonly used in OPAL analyses. It has turned out that
this matching scheme gives the best description of
data~\cite{bib-OPALresummed}, and that the results from other matching
schemes with comparable quality do not significantly differ from the
standard results.

\vspace*{1ex}
\begin{raggedright}
{\em L.~Trentadue, Parma:}  \\
\end{raggedright}
Is the quality of the data sufficiently good that the hadronisation
effects are small enough at low energies?

\vspace*{1ex}
\begin{raggedright}
{\em P.A.M~Fern\'andez:}  \\
\end{raggedright}
Of course, the hadronisation effects and its uncertainties get larger
at lower energies, but they remain small enough to perform reasonable
measurements within the fit ranges chosen in this analysis, even at
$\sqrt{s}=22$ GeV.

\vspace*{1ex}
\begin{raggedright}
{\em L.~Trentadue, Parma:} \\
\end{raggedright}
It would be useful to consider the variable
`energy-energy-correlation' for which the analysis of PETRA data has
been done at NLLA level in the early eighties.

\vspace*{1ex}
\begin{raggedright}
{\em P.A.M~Fern\'andez:} \\
\end{raggedright}
I agree, but I focussed only on the standard set of observables which
OPAL uses for $\alpha_s$ measurements at high energies, in order to
provide comparable results over a large range of the e$^+$e$^-$
continuum.

\end{document}